# Sublattice dependence and gate-tunability of midgap and resonant states induced by native dopants in Bernal-stacked bilayer graphene

F. Joucken,[1,2,*] C. Bena,[3] Z. Ge,[1] E. A. Quezada-Lopez,[1] F. Ducastelle,[4] T. Tanagushi,[5] K. Watanabe,[6] J. Velasco Jr.[1,*]

[1]*Department of Physics, University of California, Santa Cruz, California, 95064 USA*

[2]*Department of Physics, Box 871504, Arizona State University, Tempe, Arizona, 85287 USA*

[3]*Institut de Physique Théorique, Université Paris Saclay, CEA CNRS, Orme des Merisiers, 91190 Gif-sur-Yvette Cedex, France*

[4]*Laboratoire d'Etude des Microstructures, ONERA-CNRS, UMR104, Université Paris-Saclay, BP 72, 92322 Châtillon Cedex, France*

[5]*International Center for Materials Nanoarchitectonics, National Institute for Materials Science, 1-1 Namiki, Tsukuba 305-0044, Japan*

[6]*Research Center for Functional Materials, National Institute for Materials Science, 1-1 Namiki, Tsukuba 305-0044, Japan*

[*]*Email: <u>frederic.joucken@gmail.com</u>; <u>jvelasc5@ucsc.edu</u>*

**Abstract:**

The properties of semiconductors can be crucially impacted by midgap states induced by dopants, which can be native or intentionally incorporated in the crystal lattice. For Bernal-stacked bilayer graphene (BLG), which has a tunable bandgap, the existence of midgap states induced by dopants or adatoms has been investigated theoretically and observed indirectly in electron transport experiments. Here, we characterize BLG midgap states in real space, with atomic scale resolution with scanning tunneling microscopy and spectroscopy. We show that the midgap states in BLG – for which we demonstrate gate-tunability – appears when the dopant is hosted on the non-dimer sublattice sites. We further evidence the presence of narrow resonances at the onset of the high energy bands (valence or conduction, depending on the dopant type) when the dopants lie on the dimer sublattice sites. Our results are supported by tight-binding calculations that agree remarkably well with the experimental findings.

**Main text:**

Midgap states induced by dopants or defects have historically been of central importance in semiconductors physics because of their key role in the electronic transport properties and on the light absorption or emission of materials.[1,2] This is also the case for two-dimensional semiconductors, which have attracted enormous attention in the last decade. Dopant/defect induced midgap states have been shown to play a central role e.g. in the exciton formation and dynamics in two-dimensional (2D) transition metal dichalcogenides (TMD).[3–7] Several reports have also established the drastic influence of theses midgap states on the transport properties of 2D semiconducting TMD.[8–10] Scanning tunneling microscopy (STM) investigations conducted on doped 2D semiconducting TMD have revealed their spatial configuration and their electronic structure at the atomic scale.[11–15] Bernal-stacked bilayer graphene (BLG) is also a semiconducting 2D material, particularly known for its unique tunable bandgap.[16–18] This unique property makes BLG especially promising for quantum information technology as it allows designing gate-defined confinement regions such as constrictions[19] or quantum dots.[20–25] Such devices require pristine electronic response free of impurity induced midgap states to maintain quantum coherence.[26]

Impurity-induced midgap states in BLG have been investigated theoretically.[27–31] The transport properties of biased doped BLG was studied by Ferreira et al.[32] and Yuan et al.[33] who predicted a plateau in conductivity around the charge neutrality point. This plateau and its dependence on the vertical electric field was later observed experimentally with transport measurements by Stabile et al.[34] The problem of localization[30] and the Kondo effect[29] has also been investigated theoretically. More recently, the sublattice dependence of the dopants was indirectly evidenced by transport in an ultra-high vacuum environment by Katoch et al.[35] To date however, no local probe characterization of impurity states in biased BLG has been reported. Such characterization offers unique information on the local modification of the local density of states (LDOS) induced by the dopants.

We report in this letter atomic-scale characterization of dopant-induced midgap states in Bernal-stacked BLG. With data obtained on dopants with both positive and negative potential sign, we show that the dopants induce a midgap state only when sitting on the non-dimer sublattice sites. We demonstrate the tunability of the midgap states with the electrostatic backgate. We also demonstrate the existence of narrow resonant states located at the onset of the high energy bands when the dopant lies on the dimer sublattice sites. Although inaccessible in transport experiments, these higher energy states likely influence the optical properties of BLG. Our findings are supported by tight-binding-based simulations agreeing remarkably well with the experiments.

The lattice structure of Bernal-stacked BLG and our convention for sublattice labelling are shown in Fig. 1a. As McCann and Koshino, we refer to the $A_2$ and $B_1$ sublattice sites as dimer sites, because of the strong coupling between the orbitals of atoms located on them.[18] By the same token, we refer to $A_1$ and $B_2$ as non-dimer sites. Figure 1b shows the band structure of Bernal-stacked BLG around the K point when an interlayer potential of $U = 100$ meV (see Fig. 1a for the polarity; other tight-binding parameters are $\gamma_0 = 3.3$ eV, $\gamma_1 = 0.42$ eV, and $\gamma_3 = -0.3$ eV;[36] following the convention from ref.[18]). Note that for simplicity we ignore the difference between the interlayer potential ($U$) and the gap ($U_G$) because it is negligible in the range we consider here ($< 90$ meV). The LDOS on each sublattice is shown in Fig. 1c. The color of each band in Fig. 1b reflects the distribution of the LDOS on each sublattice, as seen in Fig. 1c. The LDOS plotted in Fig. 1c show the strong lattice dependence of the electronic density in gapped BLG. The strong van Hove singularities (vHs) at the gap edges (low energy) are spatially located around the non-dimer sites, whereas the weaker vHs corresponding to the onset of the higher energy bands can be associated with the dimer sites.

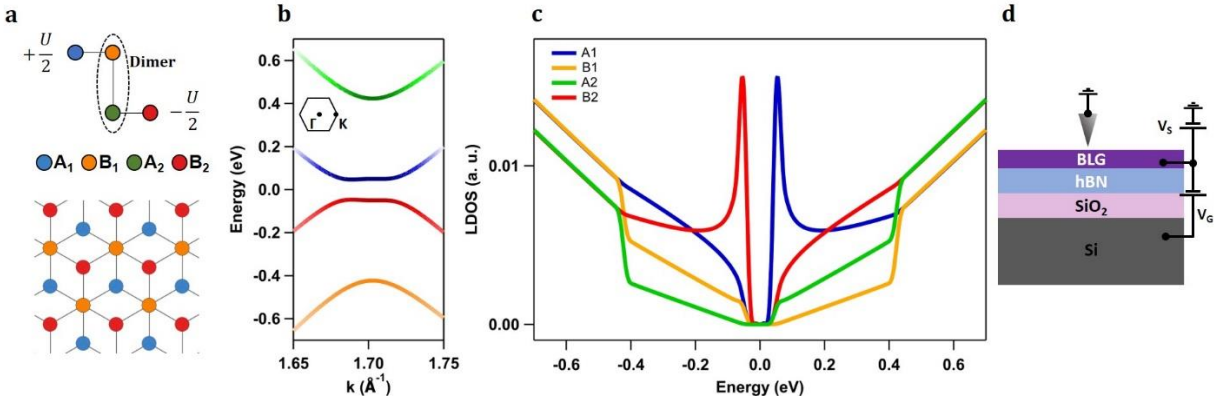

Fig. 1: **Bernal-stacked bilayer graphene (BLG) lattice, band structure, and local density of states (LDOS).** (a) Lattice structure of Bernal-stacked BLG. (b) Band structure of Bernal-stacked BLG, around the K point. (c) Local density of states on each sublattice site for U=+100 meV. The color code used in panel (b) is indicative of the LDOS on each sublattice site. (d) Schematics of the backgated Bernal-stacked BLG device investigated by scanning tunneling microscopy in this work.

The sample we have investigated consists in a Bernal-stacked bilayer graphene (BLG)/hBN heterostructure deposited on $SiO_2$/Si (see SM section 1, which includes refs.[24,37–39], for details on sample fabrication). In our STM experiments the tip is grounded, and a bias is applied to the sample ($V_G$) while the silicon substrate serves as an electrostatic gate ($V_G$); a schematic of the device is shown in Fig. 1d. As we have recently reported,[40] native defects/dopants are found in these samples. It is challenging to find and characterize these dopants with STM because, although their effects are striking,[40] their concentration is extremely low ($\sim 1\,2 \times 10^9$ cm$^{-2}$). However, an advantage of having such a low concentration of defects/dopants is that interactions between them[41,42] are absent, simplifying somewhat the comparison between experiments and simulations.

We start by presenting the sublattice dependence of the dopant-induced localized states in gapped Bernal-stacked BLG (Fig. 2). Figures 2a and 2b show a dI/dV$_S$ spectra acquired atop the dopants shown in the STM topography images displayed in the respective insets (red crosshair).

The dotted gray spectra in Fig. 2a and 2b are reference spectra obtained ∼3 nm away from the dopants. The spectra were obtained at $V_G = +30$ V, corresponding to $U_G = 64$ meV (see SM section 2, which includes refs.[43–45], for details on the band gap determination). Superimposing atomic lattice schematics to the STM image (see SM section 3, which includes ref.[46]) suggests that the dopant in Fig. 2a is located on the $B_1$ sublattice (dimer site) whereas the dopant in Fig. 2b is located on the $A_1$ sublattice (non-dimer site). The spectra of Fig. 2a and 2b both display a broad state in the conduction band ($V_S \approx +0.75$ V) similar to what is observed atop nitrogen dopant in monolayer graphene,[47] indicating the dopant should be donor-like, with attractive potential.[48] We refer to this state as the *broad resonance*.

The spectroscopic and the topographic STM signatures of the dopants shown in Figs. 2a and 2b are very similar to the signature of nitrogen dopants in monolayer graphene.[47,49,50] Also, the topographic STM signature of nitrogen dopant in bilayer graphene was shown to resemble closely the signature in monolayer graphene.[51,52] We thus identify the dopants as nitrogen. The spectra in Fig. 2a and 2b also display two very clear differences. The first difference is found close to the Fermi level ($V_S = 0$): a state is clearly visible for the dopant on the $A_1$ sublattice, while no state is seen for the dopant on the $B_1$ sublattice. It lies at the conduction band gap edge (see below). We refer to it as the *midgap state* and associate it to the predictions made by earlier theoretical work,[27,31] as we show below.[1] Another difference can be seen at $V_S \approx 0.40$ V, where a narrow peak in the spectrum can be seen only for the defect located on the $B_1$ sublattice, no such state being seen for the defect on $A_1$. We refer to this state as the *narrow resonance*. Importantly, these three states (midgap state, broad and narrow resonances) are well localized on the dopants, as they are not present ∼ 3 nm away from the dopants (gray spectra in Fig. 2a and 2b; we also present more spatially dependent spectra in the SM (section 4). Finally, note that in pristine regions, the sharp peak associated with the low energy vHs, which could be expected to appear in the STS spectra (red and blue curves in Fig. 1c), turns out to be very weak. Also, over the energy range that we have probed (+/- 1 eV), only a faint sublattice dependence is observed in the pristine regions.[53] We discuss this point in detail in section 5 of the SM, which includes ref.[53].

---

[1] Note that although the state lies at the edge of the gap, we follow the terminology of Nilsson and Castro Neto[27] and refer to it as the *midgap* state.

To clarify the origin of these states, we next compare our experimental data to tight-binding simulations. Figures 2c and 2d display computed LDOS on $B_1$ and $A_1$ (respectively) for a dopant located on the $B_1$ and $A_1$ (respectively). The dopant is modelled by modifying the onsite energy at the specific sublattice position and the results are shown for a range of onsite energy values (details on the calculations are given in section 6 of the SM, which includes refs.[54,55]). A very good agreement between the simulated LDOS and the experimental $dI/dV_S$ spectra can be seen. In both cases, the broad resonance in the conduction band is reproduced by the simulations and the position in energy is comparable to experiment. Notably, the state next to the Fermi level is also well reproduced in the simulations. In addition, the narrow resonance at ≈ 0.40 eV for the $B_1$ case is well reproduced. Note that the position of the state in energy coincides with the onset of the high energy band (Fig. 1b) and that simulations show that a dopant located on $A_2$ (the other dimer site) results in very similar LDOS to a dopant located on $B_1$ (peak in the LDOS also located in the conduction band; see supp. mat.). This narrow resonance located at the high energy band onset can be expected because of the van Hove singularity (discontinuity of the density of states and divergence of its Hilbert transform) associated with the dimer sites ($B_1$ and $A_2$ sublattices; see Fig. 1). As the midgap state (edge of the low energy band) is produced by the dopant lying on a non-dimer site, a dopant lying on a dimer site can be expected to produce a resonant state located at the edge of the associated high energy band (see Fig. 1c).[56]

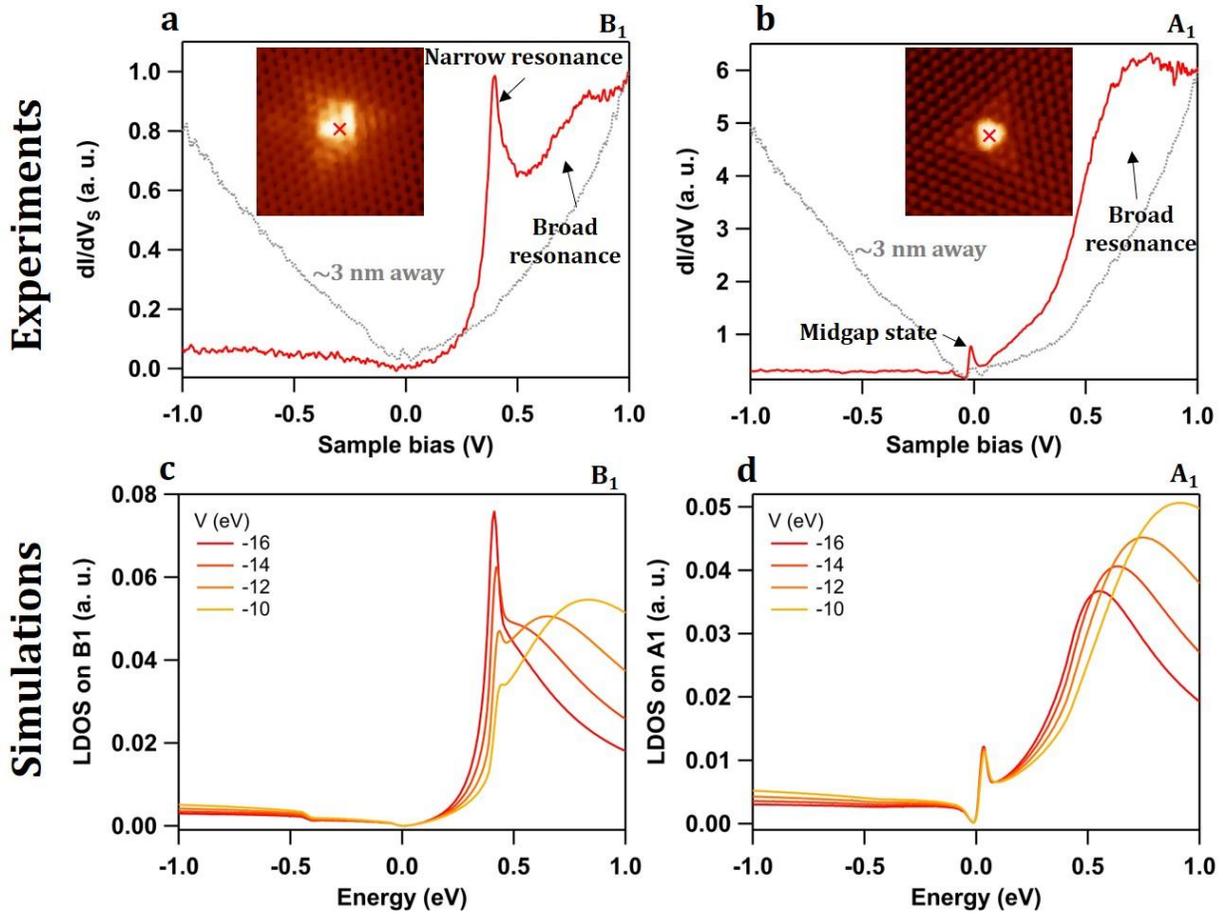

*Fig. 2: **Sublattice dependence of the spectroscopic signature of a nitrogen dopant in Bernal-stacked BLG.** (a) Experimental $dI/dV_S$ spectrum acquired atop a nitrogen dopant located on the $B_1$ sublattice. Inset: 3×3 nm² topographic STM image of the dopant; z-scale is 0.14nm, $V_S$=+1V, I=0.5nA, $V_G$=+30V. (b) Experimental $dI/dV_S$ spectrum acquired atop a nitrogen dopant located on the $A_1$ sublattic. Inset: 3×3 nm² topographic STM image of the dopant; z-scale is 0.14nm, $V_S$=-0.3V, I=0.5nA, $V_G$=+30V. Data acquired at a gate voltage $V_G = +30$ V and stabilizing parameters $I = 0.5$ nA, $V_S = -1$ V. Spectra were acquired at the location indicated by the red crosshair in the topographic image. In (a) and (b), the grey spectrum is a reference spectrum acquired in a pristine region, ~3 nm away from the dopant, in the same conditions, for comparison. (c) Simulated LDOS of a dopant located on the $B_1$ sublattice, modeled by the indicated onsite energy ($V = -10, -12, -14, and -16$ eV) (d) Simulated LDOS of a dopant located on the $A_1$ sublattice, modeled by the indicated onsite energy ($V = -10, -12, -14, and -16$ eV). The main features seen in (a) and (b) are well reproduced in (c) and (d).*

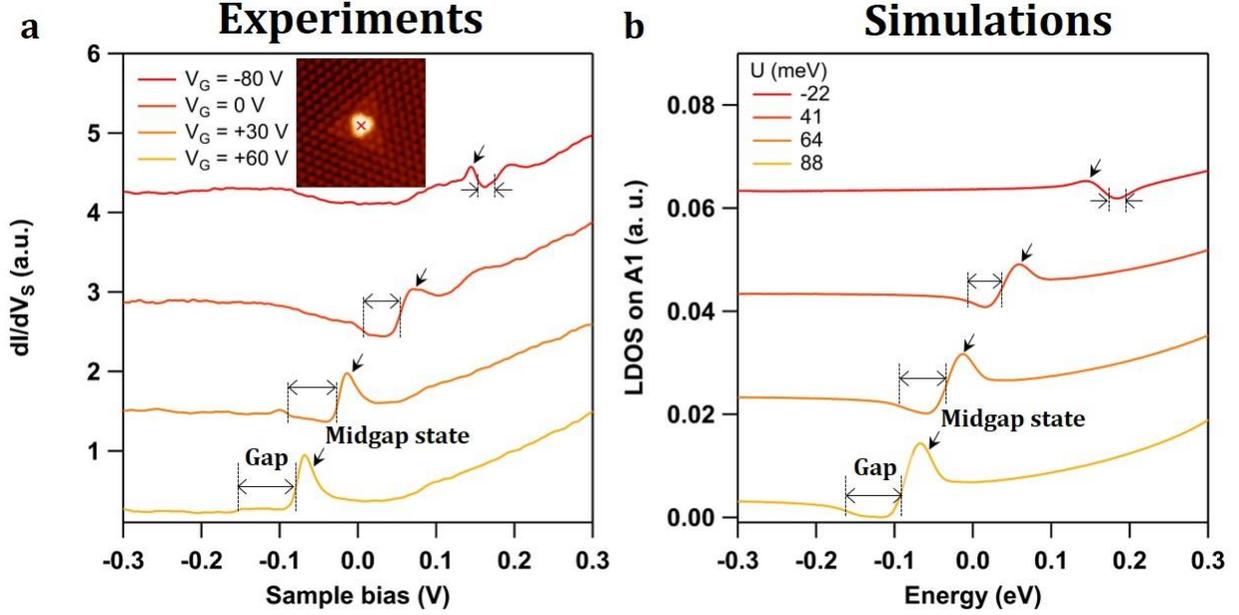

*Fig. 3: **Gate dependence of the midgap state associated to a nitrogen dopant in Bernal-stacked BLG.** (a) Experimental $dI/dV_S$ spectra obtained atop a nitrogen dopant located on the $A_1$ sublattice for various gate voltages. Inset: : 3×3 nm² topographic STM image of the dopant; z-scale is 0.14 nm, $V_S = -0.3$ V, $I = 0.5$ nA, $V_G = +30$ V. The spectra were acquired at the position indicated by the red crosshair in the topographic image. (b) Simulated LDOS for the $A_1$ sublattice for a dopant modelled by an onsite energy $V = -12$ eV. In (a) and (b) the midgap state and the gap are indicated. For unknown reason, the gap edge on the valence band side is not clearly visible for $V_G = +60$ V in (a). Curves in (a) and (b) are shifted vertically for clarity. The simulated LDOS are shifted manually in energy to match the position of the midgap state in the experiments.*

We now explore the gate tunability of the midgap state. In Fig. 3a we present the gate-dependent STS data obtained atop the nitrogen dopant located on the $A_1$ sublattice (same dopant as in Fig. 2b). The simulated LDOS for the corresponding gap values (see SM section 2 for gap determination) is shown in Fig. 3b, for an onsite potential $V = -12$ eV. The LDOS curves in Fig. 3b have been shifted vertically for clarity and horizontally to match the gate dependence observed experimentally (Fig. 3a). The gap is indicated on the LDOS curves in (b), as well as in the experimental data in (a). The gap edge on the valence band side is not clearly resolved experimentally for $V_G = +60$ V in (a) (unknown reason). A very good agreement between the simulations and the experiment is evident. As the gate (and the gap) increases, the midgap state becomes more prominent, as can be intuitively expected since the midgap state disappears when the gap closes.[27,31] Note also that we observe the midgap state very close to the conduction band edge. The separation between the midgap state and the band edge is not resolved within our experimental resolution, which we estimate at 15 meV (the amplitude of the lockin excitation; see section 1 of SM). This is in agreement with the values anticipated theoretically.[27,31] We must however note that, although the state seen experimentally in Fig. 3a can be unequivocally attributed to the midgap state, the simulated LDOS curves for the pristine case (in the absence of defect) display a similar shape as seen in Fig. 3b, only with a less intense peak at the band edge (see section 5 of the SM). That peak corresponds to the low energy vHs expected on $A_1$ (see Fig. 1). As already mentioned, the low energy vHs peak in pristine regions turns out to be very weak in STS spectra.[53]

In addition to the dopant that causes the midgap state discussed so far we observed another type of dopant that also produces a broad resonance, but in the valence band. Figure 4a shows dI/dV$_S$ spectra obtained atop the dopants whose STM topographic images are shown as insets. We also reproduce in Fig. 4a the experimental spectrum obtained atop the N-dopant located on $A_1$ in light dotted grey (Fig. 2b), for convenient comparison. Both dopants display a broad resonance at $V_S \approx -0.75$ V. However, only the red curve displays a midgap state (narrow peak close to $V_S = 0$). Also, the gold curve shows a slight shoulder around $V_S = -0.4$ V.

To understand the origin of the states seen experimentally (Fig. 4a), we show in Fig. 4b simulated LDOS for dopants modelled by an onsite potential $V = +12$ eV located on $A_2$, $B_2$, and $B_1$ (we also show the LDOS for $V = -12$ eV on $A_1$, for comparison). Because of the agreement between the red solid line curves in Figs. 4a and 4b, we can confidently assign the red spectrum in Fig. 4a to a dopant located on the $B_2$ sublattice (non-dimer site) with repulsive potential ($V = +12$ eV). For the other unknown dopant (gold curve) the situation is less straightforward as both simulated LDOS for a dopant on the $B_1$ and the $A_2$ sublattices (dimer sites) with $V = +12$ eV reproduce well the experimental dI/dV$_S$ spectrum: the absence of midgap state and the shoulder at the onset of the high energy band ($V_S = -0.4$ V) are both reproduced. Note that $V_S = -0.4$ V corresponds to the onset of the high energy band (similarly as the nitrogen dopant located on the $B_1$ sublattice, see Fig. 2a). However, the similar low level of corrugation observed in the topographic STM images for both dopants indicates they both are most likely located in the bottom layer.[51] We thus tentatively assign the dopant corresponding to the gold curve to the sublattice $A_2$. Superimposed atomic lattice schematics on the STM images of the dopants are consistent with a dopant on $B_2$ (red) and $A_2/B_1$ (gold) (see SM section 3).

It is difficult to identify the chemical nature of the dopants presented in Fig. 4a but a natural candidate is boron. The symmetry between nitrogen and boron for the broad resonance in the conduction or the valence band is expected from density functional calculations[51,57] and previous tunneling spectroscopy data on monolayer graphene suggested the presence of the broad resonance in the valence band.[57–59] Boron dopants are also expected to act as negatively charged defects (as opposed to nitrogen dopants which act as positively charged defects) and we show in the SM (section 7, which includes refs.[60,61]) spatially dependent dI/dV$_S$ spectra consistent with these polarities.

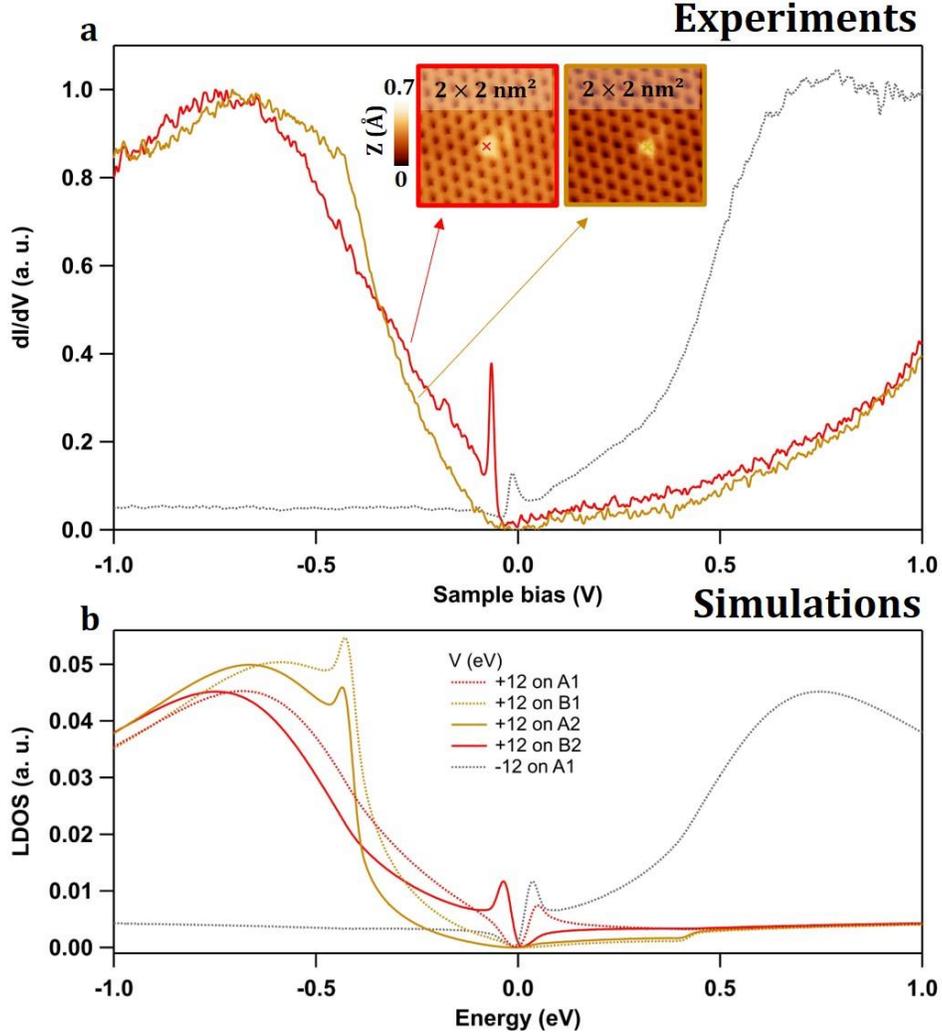

Fig. 4: **Spectroscopic signature of dopants other than nitrogen in BLG.** (a) Experimental dI/dV$_S$ spectrum obtained atop the dopants shown in the inset ($I = 0.5$ nA; $V_S = -1$ V; $V_G = +30$ V). Crosses indicate the position where the spectra were acquired. The correspondence is indicated by the arrows. The dotted grey curve is the experimental dI/dV$_S$ spectrum obtained atop the nitrogen dopant located on the A$_1$ sublattice (Fig. 2b) shown here for comparison. (b) Simulated LDOS for a dopant modelled by an onsite potential $V = +12$ eV on all sublattice sites; the grey curve is the computed LDOS at A$_1$ for an onsite potential $V = -12$ eV on A$_1$, for comparison. The location of the LDOS corresponds to the dopant site

Finally, from Fig. 4b (and Figs. 2c and d), we summarize the rules for the position of the midgap state and the narrow and broad resonances. First, the sign of the onsite potential determines the position of the broad resonance: $V > 0$ ($V < 0$) gives a broad resonance in the valence (conduction) band. Second, the narrow resonance at the onset of the high energy band is present if the dopant is on a dimer site ($B_1$ or $A_2$). Its position (valence or conduction band) does not depend on the sign of the electric field, but only on the sign of the onsite potential. Its amplitude varies significantly with the onsite potential amplitude (see Fig. 2c). Moreover, the midgap state is present if the dopant is located on a non-dimer site ($A_1$ or $B_2$). It is flanking either the valence band or the conduction band edge, depending on which dimer site it is occupying and the orientation of the electric field. The dependence on the electric field is not illustrated in Fig.4 but we show in the SM (section 8) complete results for the positions of the resonant states (broad and narrow) and the midgap state, as a function of the sublattice position, the onsite potential, and the electric field direction.

In conclusion, we have reported gate-dependent scanning tunneling spectroscopy results demonstrating the existence and the tunability of midgap states induced by native dopant in Bernal-stacked BLG. We have demonstrated that the midgap state appears when the dopant lies in the non-dimer sites and shown how it can be tuned by the backgate voltage (perpendicular electric field). We have also shown that when the dopant lies in the dimer sites, a narrow resonance appears at the edge of the high energy bands. These higher energy states, although inaccessible in transport experiments, should be observable (and tunable) for example in optical absorption spectra,[62] for samples with high enough dopants concentration, in which case subbands are expected to appear.[27,31] Studying the evolution of these states with increasing dopants concentration achieved by intentional doping with angle-resolved photoemission spectroscopy and/or STM could reveal the development of these gate-tunable impurity bands.

*We dedicate this work to the memory of our co-author François Ducastelle who passed away during the review process of this article.*

**Author contributions**

**Funding Sources**

J.V.J. acknowledges support from the National Science Foundation under award DMR-1753367 and the Army Research Office under contract W911NF-17-1-0473. K.W. and T.T. acknowledge support from the Elemental Strategy Initiative conducted by the MEXT, Japan, Grant Number JPMXP0112101001 and JSPS KAKENHI Grant Number JP20H00354.

# Supplementary Material for

# Sublattice dependence and gate-tunability of midgap and resonant states induced by native defects in Bernal-stacked bilayer graphene

*F. Joucken, C. Bena, Z. Ge, E. A. Quezada-Lopez, F. Ducastelle, T. Tanagushi, K. Watanabe, J. Velasco Jr.*

**Contents:**

1- Sample fabrication and scanning tunneling microscopy measurements
2- Extraction of the gap and the electronic doping for the BLG device for each gate voltage
3- Sublattice determination from STM images
4- Spatially-dependent STS spectra of the midgap state
5- Local density of states calculations
6- Van Hove singularity observation
7- Dependence of the midgap state, the narrow resonance, and the broad resonance on the electric field and the onsite potential
8- Spatially dependent STS spectra for determination of the sign of the dopant charge

1- Sample fabrication and scanning tunneling microscopy measurements

The bilayer graphene (BLG) was stacked on hBN using a standard polymer-based transfer method.[1] A graphene flake exfoliated (from "Flaggy Flakes" graphite from NGS Naturgraphit) on a methyl methacrylate (MMA) substrate was mechanically placed on top of a 20 nm thick hBN flake that rests on a SiO2/Si++ substrate where the oxide is 285 nm thick. Subsequent solvent baths dissolve the MMA scaffold. After the Graphene/hBN heterostructure is assembled, an electrical contact to graphene is made by thermally evaporating 7 nm of Cr and 200 nm of Au using a metallic stencil mask. The single-terminal device is then annealed in forming gas (Ar/H2) for six hours at 400 °C to reduce the amount of residual polymer left after the graphene transfer. To further clean the sample's surface, the heterostructure is mechanically cleaned using an AFM.[2,3] Finally, the heterostructure is annealed under UHV at 400 °C for several hours before being introduced into the STM chamber.

The STM measurements were conducted in ultra-high vacuum with pressures better than $1 \times 10^{-10}$ mbar at 4.8 K in a Createc LT-STM. The bias is applied to the sample with respect to the tip. The tips were electrochemically-etched tungsten tips, which were calibrated against the Shockley surface state of Au(111) prior to measurements. $dI/dV_S$ spectra were acquired with standard lockin technique, applying a 15 meV modulation to the sample bias, at 704 Hz. The STM images were treated with WSxM.[4]

2- Extraction of the gap and the electronic doping for the BLG device for each gate voltage

To extract the values of the gap in our sample for a given gate voltage, we proceeded as follows. We used a series of $dI/dV_S$ measurements at various gate voltages, presented in Fig. S1a. The horizontal lines corresponds to onset of phonon-assisted tunneling in the BLG sample.[5,6] The phonon-assisted tunneling renders the visualization and determination of the gap problematic for energies above the inelastic tunneling threshold.[7] We thus determined the gap value at gate voltage where the gap can be measured precisely (15 V), and also determined the gate voltage where the gap closes (indicated by the red arrow), and extrapolated linearly between these two values (Fig. S1b).

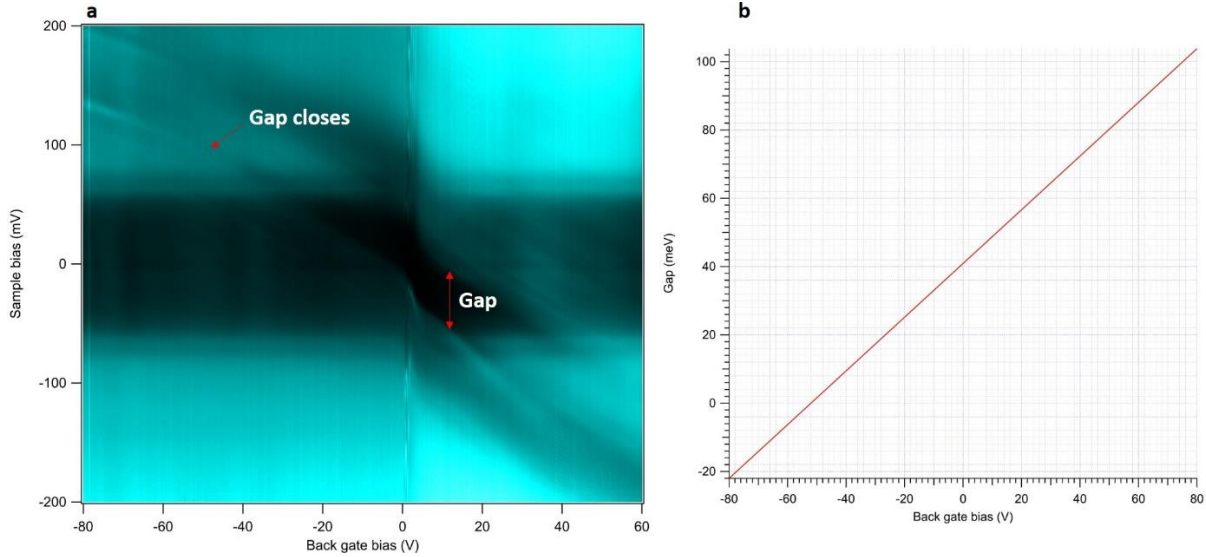

*Fig. S2 Determination of the gap for the BLG device. (a) Plot of dI/dV$_S$ spectra as a function of gate voltage for the bilayer graphene sample used in this study. The gap is clearly visible for gate voltage around $V_G = +15$ V. The gap also clearly closes around $V_G = -52$ V. (b) Gap values linearly extrapolated from (a).*

3- <u>Sublattice determination from STM images</u>

We show in Fig. S2 the superposition of BLG lattice on the STM images shown in the main text and for the dopant shown in Fig. S3. Note that for the dopant shown in Fig. S2a (A$_1$), the superimposition does not provide conclusive agreement. However, the overall shape of triangular patterns associated with the dopants are clearly rotated by 60° between the two cases, providing clear indication that the two dopants sit on different sublattices.[8]

4- <u>Spatially-dependent STS spectra of the midgap state</u>

Figure S3 displays dI/dV$_S$ spectra acquired along the black arrow (~6 Å step size) on the STM topographic image of the dopant lying on the A$_1$ sublattice shown in the inset (it is a different dopant than in Fig. 2b). It is evident from the data that the midgap state is observed atop the dopant and vanishes already ~6 Å away from it. Small peaks can be seen further away at the same energy position as the midgap state; we attribute these to the expected van Hove singularity which is sublattice dependent (see Fig. 1 of the main text). The dI/dV$_S$ spectra acquired along the line shown in Fig. 3 can sometimes be acquired closer to the A$_1$ sublattice, sometimes closer to the B$_1$ sublattice). We discuss this point further in the section 6 below.

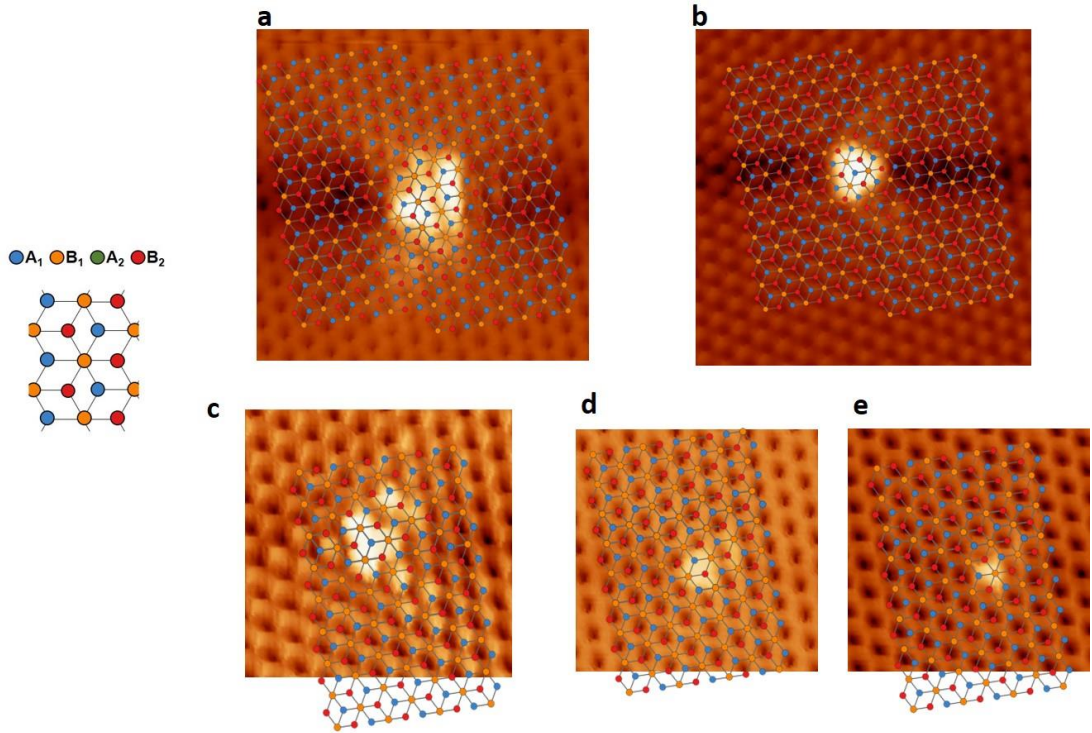

*Fig. S3 Superposition of BLG lattice on STM images for the dopants shown in Fig. 2a and 3 (a), Fig. 2b (b), Fig. S3 (c), and Fig. 4 (d and e) of the main text. The dopants are located on $B_1$ (a), $A_1$ (b), $A_1$ (c), $B_2$ (d), and $A_2$ (e). Note that for (a) ($B_1$), the superimposition does not provide conclusive agreement. However, the overall shape of triangular patterns associated with the dopants are clearly rotated by 60° compared to (b) ($A_1$), providing clear indication that the two dopants sit on different sublattices.*

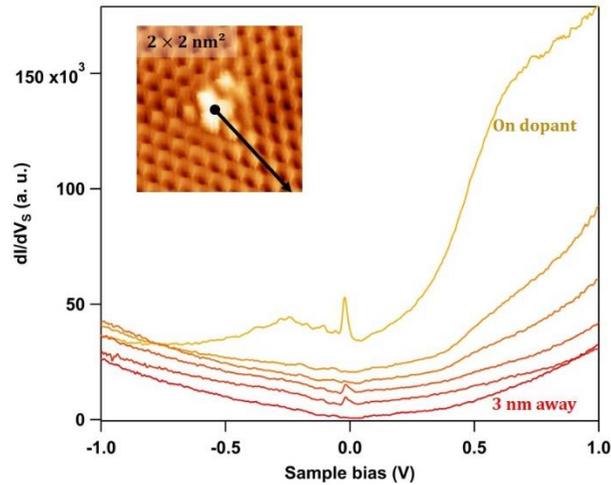

*Fig S4 Experimental dI/dV spectra acquired along the black arrow for a nitrogen dopant sitting on the $A_1$ sublattice (STM topographic image shown in the inset). ($I = 0.5$ nA, $V_S = -1$ V, $V_G =$*

+30 *V.) The midgap state, as well as the broad resonance in the conduction band, rapidly vanish as one moves away from the dopant.*

5- <u>van Hove singularity observation</u>

We mention in the main text that the van Hove singularity (vHs) is barely seen with STM in the pristine case. We show in Fig. S4a and S4b dI/dV$_S$ spectra acquired for some values of gate voltage atop the A$_1$ and B$_1$ sublattices (respectively), in a pristine region. One can see that no vHs is observed for $U > 0$, neither atop A$_1$ nor B$_1$. Only for $U = -22$ meV one can see a peak flanking the valence band edge, corresponding to the expected vHs (see Fig. 1 of main text). We show in Fig. S4c and S4d more dI/dV$_S$ spectra (arranged in a 2D map as a function of V$_S$ and V$_G$) confirming this observation: the vHs is observed with STS only for gate voltage values smaller than the gate voltage value for which the gap closes ($V_G \approx -52$ V). It is more pronounced when measured atop A$_1$ (Fig. S4c), but still present atop B$_1$ (Fig. S4d). The reason for this asymmetry observed experimentally is unknown to us but we have observed it consistently, with many different STM tips. Further data and discussion on this issue can be found in E. A. Quezada-Lopez PhD dissertation.[9]

We mention in the main text that the computed LDOS on A$_1$ in the pristine case and in case a dopant is lying on A$_1$ have similar shape. We present in Fig. S4 LDOS on A$_1$ for these cases. Figures S4 a, S4b, and S4c compare the intensity of the peaks (midgap state or vHs, depending on whether we consider the doped or the pristine case, respectively) relative to the LDOS at $-0.2$ eV for various values of gap (indicated). One can see that the prominence of the midgap state increases with increasing gap values. That is consistent with the experimental findings of Fig. 4a in the main text.

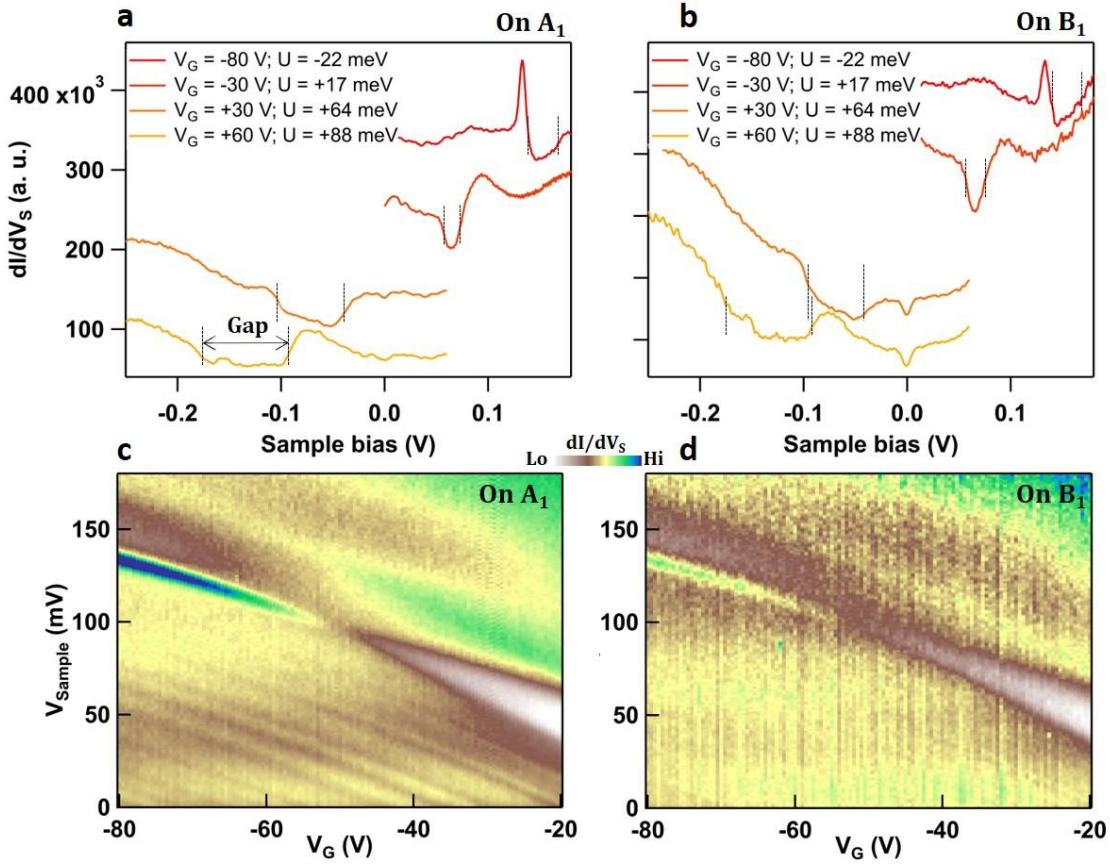

*Figure S5: dI/dV$_S$ spectra acquired for some values of gate voltage atop the A$_1$ (a) and B$_1$ (b) sublattices, in a pristine region. One can see that no vHs is observed for U > 0, neither atop A$_1$ nor B$_1$. Only for U = −22 meV a peak flanking the valence band edge is present, corresponding to the expected vHs (see Fig. 1 of main text). (c) and (d) are more dI/dV$_S$ spectra (arranged in a 2D map as a function of V$_S$ and V$_G$) obtained atop the A$_1$ (c) and B$_1$ (d) sublattices confirming this observation: the vHs is observed with STS only for gate voltage values smaller than the gate voltage value for which the gap closes (V$_G$ ≈ −52 V). It is more pronounced when measured atop A$_1$ (c), but still present atop B$_1$ (d).*

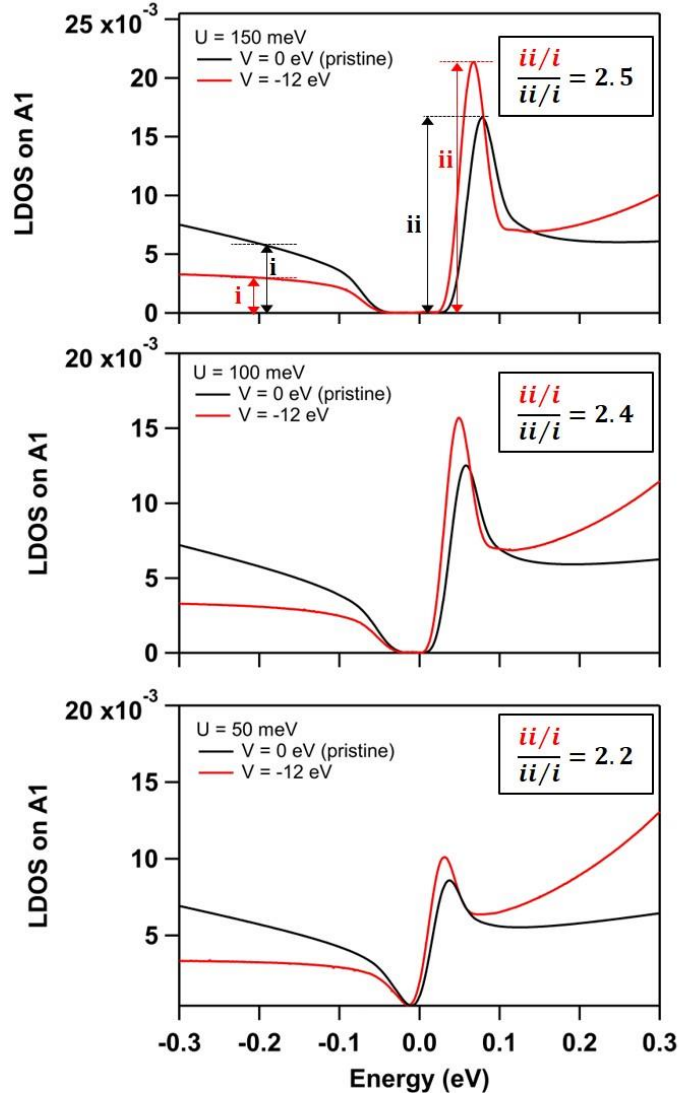

*Fig. S6: LDOS on $A_1$ for the pristine case and for an onsite potential of -12eV for various values of the gap (indicated). We compare the intensity of the peaks (midgap state or vHs, depending on whether we consider the doped or the pristine case, respectively) relative to the LDOS at -0.2eV for all cases. One can see that the prominence of the midgap state increases with increasing gap values. That is consistent with the experimental findings of Fig. 4a in the main text*

6- <u>Local density of states calculations</u>

To compute the local density of states presented in the main text, we used the pybinding package.[10] We used a circular bilayer graphene area with a radius of 100 nm with the dopant at the center. The dopant is modelled by changing the onsite energy of the sublattice site where it is lying (the neighbors are unchanged). The computation used the kernel polynomial method,[11] as implemented in pybinding. A broadening width of 15 meV was used, unless stated otherwise.

7- <u>Spatially dependent STS spectra for determination of the sign of the dopant charge</u>

In Fig. S7, we show spatially-dependent $dI/dV_S$ spectra acquired in the vicinity of N and B dopants (indicated). We follow the methodology presented in ref.[12] and ref.[13] to determine the sign of the charge associated to each dopant. All spectra are normalized by their value at $V_S = -1V$. For N dopants, the intensity of the $dI/dV_S$ signal is significantly greater on the conduction band side when you approach the dopants. This asymmetry can be explained by a shift of the Dirac point corresponding to greater electron doping as you approach the dopant. On the contrary, for B dopants, the intensity of the $dI/dV_S$ signal is smaller on the conduction band side when you approach the dopant. This can be explained by a shift of the Dirac point corresponding to smaller electron doping as you approach the dopant. That is consistent with N dopant being charged positively (attracting electrons) and B dopants being charged negatively (attracting holes). Note the effect is less strong for B dopants. We attribute this to the fact that they are located in the bottom layer.

8- <u>Dependence of the midgap state, the narrow resonance, and the broad resonance on the electric field and the onsite potential</u>

LDOS on $A_1$, $B_1$, $A_2$, $B_2$ for a dopant modelled by an onsite potential on the corresponding site for various values of the electric field and the onsite potential is shown in Fig S5. The broad resonance position (valence or conduction band) is determined by the onsite potential sign. The narrow resonance is present only when the dopant lies in a dimer site ($B_1$ or $A_2$). Whether it is located at the edge of the high energy valence or conduction band depends only on the sign of the onsite potential. The midgap state is present only when the dopant is lying on a non-dimer site ($A_1$ or $B_2$). Whether it is flanking the conduction or the valence band edge depends on the electric field sign.

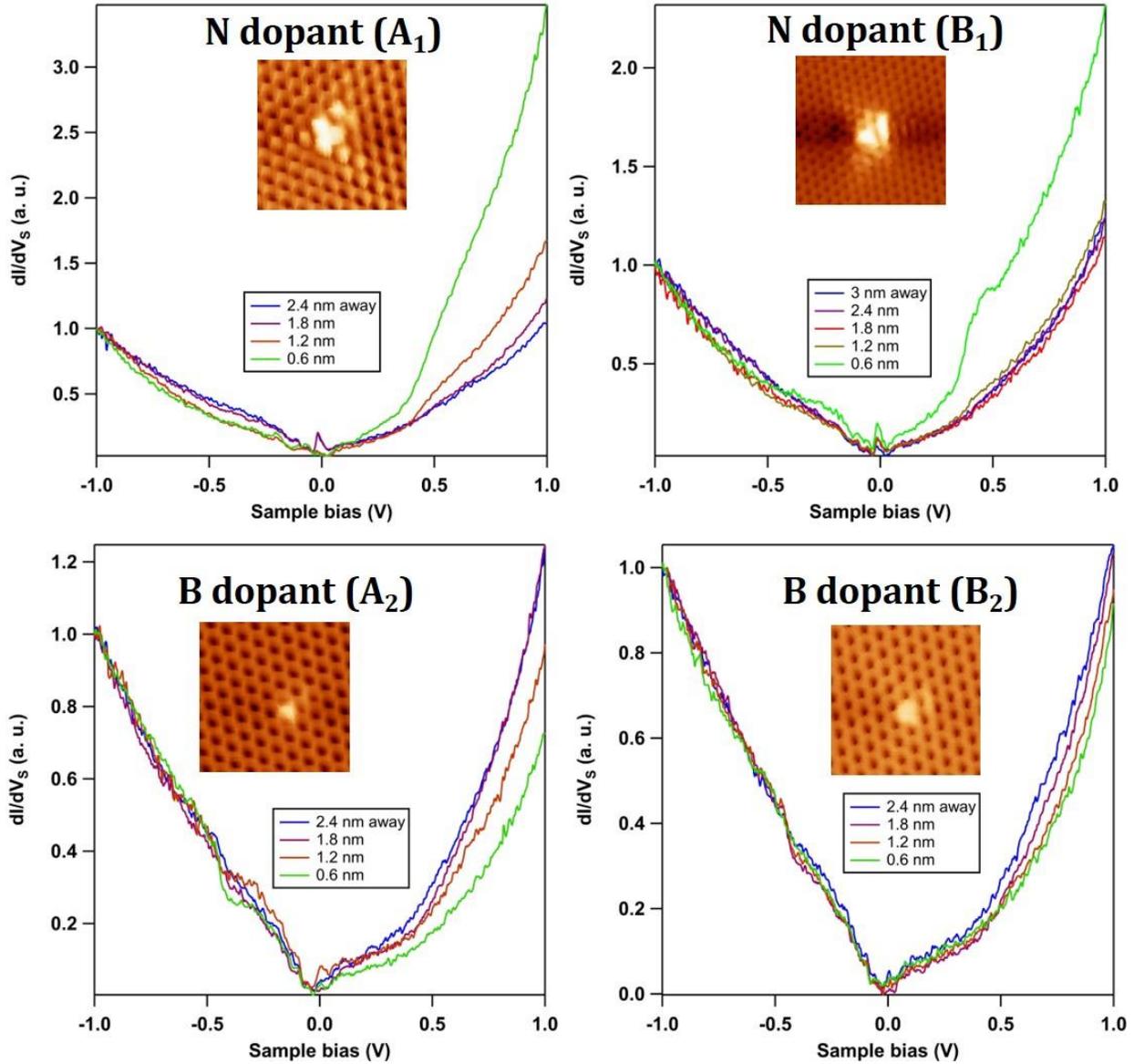

*Fig. S7: Spatially-dependent $dI/dV_S$ spectra acquired in the vicinity of dopants. The type of dopant (N or B) and sublattice locations are indicated. All spectra are normalized by their value at a sample bias of -1V. For N dopants, the intensity of the $dI/dV_S$ signal is significantly greater on the conduction band side when you approach the dopants. This asymmetry can be explained by a shift of the Dirac point corresponding to greater electron doping as you approach the dopant. On the contrary, for B dopants, the intensity of the $dI/dV_S$ signal is smaller on the conduction band side when you approach the dopant. This can be explained by a shift of the Dirac point corresponding to smaller electron doping as you approach the dopant. That is consistent with N dopant being charged positively (attracting electrons) and B dopants being charged negatively (attracting holes).*

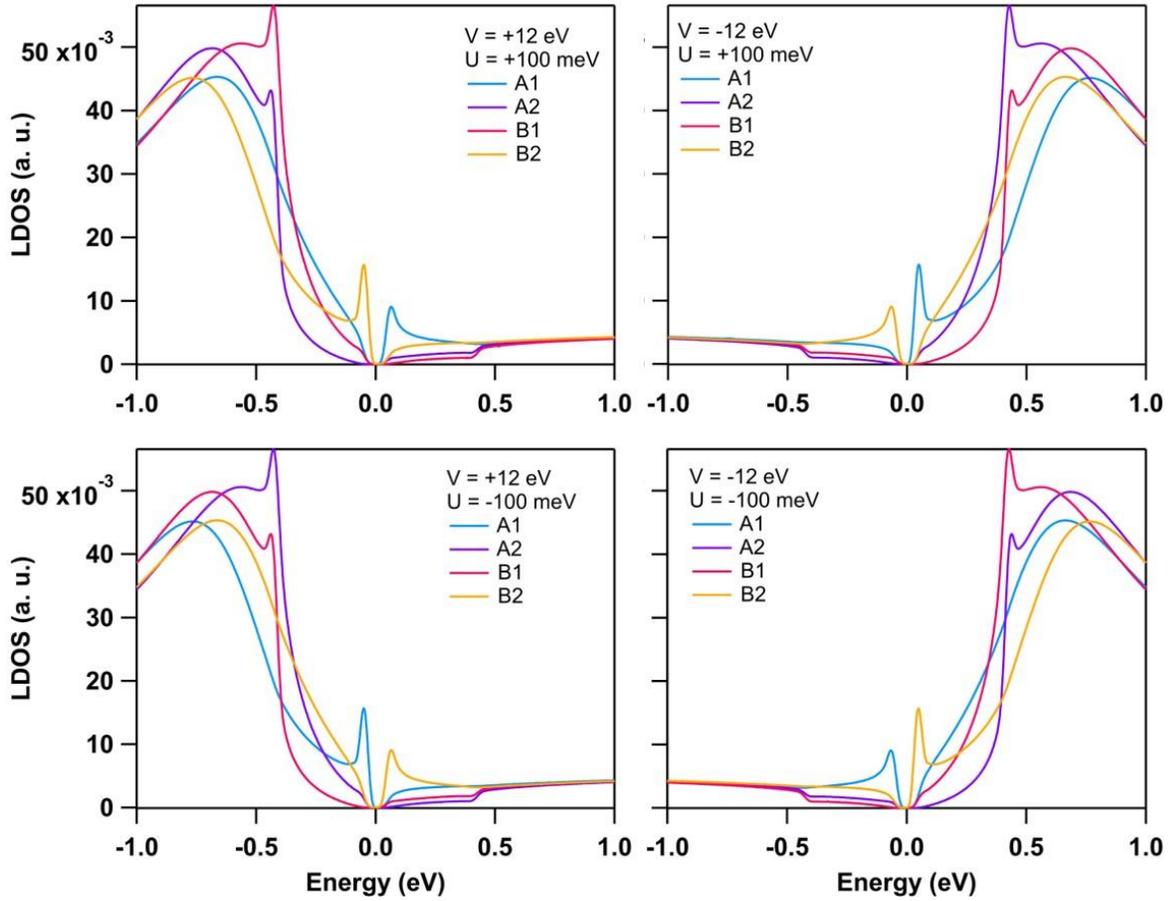

*Fig. S8: LDOS on $A_1$, $B_1$, $A_2$, $B_2$ for a dopant modelled by an onsite potential on the corresponding site for various values of the electric field and the onsite potential (indicated). The broad resonance position (valence or conduction band) is determined by the onsite potential sign. The narrow resonance is present only when the dopant lies in a dimer site ($B_1$ or $A_2$). Whether it is located at the edge of the high energy valence or conduction band depends only on the sign of the onsite potential. The midgap state is present only when the dopant is lying on a non-dimer site ($A_1$ or $B_2$). Whether it is flanking the conduction or the valence band edge depends on the electric field.*